\documentclass[pra,aps,twocolumn,10pt,showpacs,groupedaddress,superscriptaddress,floatfix,notitlepage]{revtex4-2}
\usepackage{amsmath,amsfonts,amssymb,graphics,graphicx,epsfig,color,times, braket}%,bbm}
\usepackage[utf8x]{inputenc}
\usepackage{color}
\usepackage{bbm, dsfont}
\usepackage{hyperref} 
\usepackage{mathrsfs}
\usepackage{verbatim} 
\begin{document}

\title{Quantum Mpemba effect in a driven-dissipative atom-waveguide coupled system}

\author{Hao-Kai Chi}
\email{b11901133@ntu.edu.tw}
\affiliation{Department of Electrical Engineering, National Taiwan University, Taipei 106319, Taiwan}
\affiliation{Institute of Atomic and Molecular Sciences, Academia Sinica, Taipei 106319, Taiwan}
\author{H. H. Jen}
\email{sappyjen@gmail.com}
\affiliation{Institute of Atomic and Molecular Sciences, Academia Sinica, Taipei 106319, Taiwan}
\affiliation{Department of Physics, National Central University, Taoyuan City 320317, Taiwan} 
\affiliation{Molecular Science and Technology Program, Taiwan International Graduate Program, Academia Sinica, Taiwan} 
\affiliation{Physics Division, National Center for Theoretical Sciences, Taipei 106319, Taiwan}
\date{\today}
\renewcommand{\r}{\mathbf{r}}
\newcommand{\f}{\mathbf{f}}
\renewcommand{\k}{\mathbf{k}}
\def\p{\mathbf{p}}
\def\q{\mathbf{q}}
\def\bea{\begin{eqnarray}}
\def\eea{\end{eqnarray}}
\def\ba{\begin{array}}
\def\ea{\end{array}}
\def\bdm{\begin{displaymath}}
\def\edm{\end{displaymath}}
\def\red{\color{red}}
\pacs{}
\begin{abstract}
	The quantum Mpemba effect (QME) describes the counterintuitive phenomenon in which a more strongly perturbed quantum system relaxes toward its steady state faster than a less perturbed one. Here, we investigate the QME in a weakly driven-dissipative atomic array coupled to a one-dimensional chiral waveguide. Starting from the steady state, we introduce a local quench by resetting a single atom to its ground state and examine the subsequent relaxation dynamics. We find that quenching atoms near the array boundaries induces faster initial relaxation through enhanced coupling to superradiant modes and direct photon emission into the waveguide, giving rise to an edge-enhanced QME. Strong directional decay further amplifies this effect by promoting the spatial accumulation of superradiant modes near the preferred boundary, leading to a skin-enhanced QME. At later times, relaxation is increasingly governed by long-lived subradiant modes localized predominantly in the bulk of the array. By coupling two atomic arrays with different interparticle spacings, we further demonstrate an interface-induced QME, where quenching an atom near the interface makes the system relax more slowly at early times but approach its steady state faster at later times, owing to weak participation in both superradiant and subradiant modes. Our findings reveal how the spatial structure of collective decay modes and directional dissipation jointly govern anomalous quantum relaxation, offering potential routes toward accelerated steady-state preparation in waveguide QED systems.
\end{abstract}
\maketitle
%%%%%%%%%%%%%%
\section{Introduction}

The Mpemba effect (ME) refers to the counterintuitive phenomenon that an initially hotter system cools faster than a colder one \cite{Mpemba1969, Carollo2021, Chatterjee2023, Chatterjee2024, Chalas2024, Rylands2024, Joshi2024, Shapira2024, Turkeshi2025, Ares2025, Zhang2025, Strachan2025, Nava2025, Ma2025,  Teza2026, Longhi2026_1, Longhi2026_2}. This anomalous relaxation behavior results from smaller overlaps with the slowest decaying eigenmode \cite{Teza2026, Zhang2025, Ares2025}, leading to an accelerated state equilibrating process which depends critically on its initial conditions. Recent advances involve quantum strong ME in a single ion by preparing an optimal initial state with vanishing overlaps with the slowest decaying mode \cite{Carollo2021, Zhang2025}, rapid relaxation process by eliminating initial coherence among degenerate slowest modes \cite{Zhan2026}, and a resource-theoretical perspective on ME \cite{Summer2025}. These studies offer a unique advantage of controlling initial mode overlaps to accelerate thermal equilibrium or relaxation process in driven-dissipative quantum systems.

The quantum Mpemba effect (QME) \cite{Xu2026, Calabrese2026} has attracted growing interest as an anomalous relaxation phenomenon in quantum systems, where a state initially farther from equilibrium can approach its stationary state faster than a closer one \cite{Ares2025}. Quantum relaxation exhibits fundamentally different characteristics in closed and open systems. In closed quantum systems, unitary evolution preserves the global purity of the state, while local equilibration can emerge through the redistribution of quantum correlations and entanglement \cite{Calabrese2026}. In this context, the QME has been associated with the dynamical restoration of symmetries, as characterized by entanglement asymmetry \cite{Calabrese2026, Joshi2024, Turkeshi2025}. In contrast, open quantum systems exchange energy and information with their environments, giving rise to irreversible relaxation toward stationary states. Such dynamics are commonly described by Lindblad master equations, where the interplay between coherent interactions and dissipative processes can generate a rich variety of relaxation behaviors \cite{Nava2024, Wang2024, Qian2025, Liu2026}. In particular, collective dissipation in quantum optical systems can produce superradiant and subradiant modes with widely separated decay timescales, leading to strongly nonuniform relaxation dynamics \cite{Harrington2022, Hotter2025, Feyisa2026}. These features provide a promising setting for investigating anomalous relaxation phenomena and their dependence on spatially resolved excitation and dissipation.

In this work, we investigate the QME in a weakly driven-dissipative atomic array coupled to a one-dimensional chiral waveguide \cite{Mitsch2014, Douglas2015, Solano2017, Lodahl2017, Chang2018, Mahmoodian2018, Mahmoodian2020, Jen2020_subradiance, Jen2020_steady, Sheremet2023, Jen2025_review, Tudela2024, Pichler2015}, following a local quench of the steady state in which a single atom is reset to its ground state. We find that atoms near the array boundaries exhibit faster initial relaxation, owing to their stronger coupling to rapidly decaying superradiant modes and the associated photon emission into the waveguide. This boundary-dependent relaxation becomes particularly pronounced under strongly directional decay, which leads to the spatial accumulation of superradiant modes near one boundary. At later times, the relaxation dynamics are instead dominated by the most subradiant mode, whose spatial profile is predominantly localized in the bulk of the array. Furthermore, by connecting two atomic arrays with different interparticle spacings through a common waveguide \cite{Chung2024, Handayana2024}, we demonstrate that the resulting interface can exhibit accelerated late-time relaxation compared with the bulk regions, owing to its reduced participation in long-lived subradiant modes.

The remainder of this paper is organized as follows. In Sec.~II, we introduce the theoretical model of the driven-dissipative atomic array coupled to a waveguide. In Sec.~III, we examine the QME associated with rapidly decaying modes near the array boundaries. In Sec.~IV, we investigate the enhancement of the QME under strongly directional decay, emphasizing the role of the non-Hermitian skin effect and the resulting spatial localization of superradiant modes. In Sec.~V, we demonstrate the emergence of the QME at the interface between two atomic arrays with different parameters. Finally, in Sec.~VI, we summarize our findings and discuss their implications.

\section{Theoretical model}

In this section, we introduce the effective Hamiltonian describing the atom–waveguide coupled system and its dissipative dynamics governed by the Lindblad master equation. We then employ the trace distance to characterize the relaxation of the system toward its steady state and identify the QME following local quenches at different atomic sites. To elucidate the underlying mechanisms of the QME, we analyze the eigenspectrum and spatial profiles of the collective decay modes, revealing how their distinct relaxation timescales and spatial distributions govern the system's relaxation dynamics depending on the location of the quenched atom. 
   
\begin{figure}[t]
	\centering
	\includegraphics{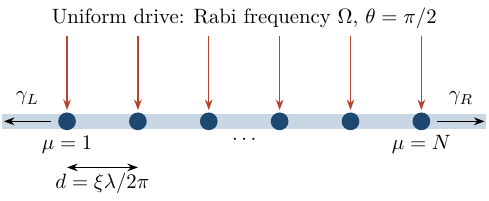}
	\caption{Schematic of the system. A chain of $N$ atoms with equal interparticle spacing $d$ is coupled to a waveguide with nonreciprocal decay rates $\gamma_\mathrm{L}$ and $\gamma_\mathrm{R}\neq \gamma_\mathrm{L}$ into opposite propagation directions and is driven uniformly from the side with a Rabi frequency $\Omega$.}\label{Schematic}
\end{figure}

\subsection{Evolution of the system}

We consider a driven-dissipative atomic array \cite{Chung2024, Handayana2024, Yu2026} as shown in Fig. \ref{Schematic}, consisting of \textit{N} two-level atoms coupled to a one-dimensional waveguide and driven by a uniform constant driving field with Rabi frequency $\Omega$ and detuning $\delta$. Under the Born-Markov approximation \cite{Lehmberg1970, Jen2025_review}, the time evolution of the density matrix $\rho$ is governed by 
\begin{equation}
\frac{d\rho}{dt}
= -\frac{i}{\hbar}[H_{\mathrm{S}}+H_{\mathrm{L}}+H_{\mathrm{R}},\rho]
+\mathcal{L}_{\mathrm{L}}[\rho]
+\mathcal{L}_{\mathrm{R}}[\rho],\label{rho}
\end{equation}
where the Hamiltonian $H_{\mathrm{S}}$ describes the light-atom interaction, 
\begin{equation}
	H_{\mathrm{S}} = \hbar \sum_{\mu=1}^{N}
	\left[ \Omega\, e^{i k x_{\mu} \cos\theta}
	\left( \sigma_{\mu} + \sigma_{\mu}^{\dagger} \right)
	- \delta_{\mu}\, \sigma_{\mu}^{\dagger} \sigma_{\mu} \right],
\end{equation}
and $H_{\mathrm{L(R)}}$ describes the waveguide-mediated interactions associated with left- and right-propagating photons \cite{Sheremet2023}, respectively, 
\begin{equation}
	H_{\mathrm{L(R)}} = -i\hbar\, \frac{\gamma_{\mathrm{L(R)}}}{2}
	\sum_{\mu < (>) \nu}^{N}
	\left( e^{i k \lvert x_{\mu} - x_{\nu} \rvert}\,
	\sigma_{\mu}^{\dagger} \sigma_{\nu} - \mathrm{H.c.} \right).
\end{equation}
In the above Hamiltonian, $k$ is the wave number $2\pi/\lambda$, $x_\mu$ is the position of atom $\mu$, and $ \sigma_{\mu}^{\dagger}$ is the dipole raising operator defined as $\sigma_{\mu}^{\dagger} = |e\rangle_\mu\langle g|$. $\theta$ quantifies the incident angle of the laser excitation. In this work, we assume that the laser excitation is perpendicular to the waveguide, i.e. $\theta = \pi/2$, in a scenario of side excitations \cite{Mitsch2014}, under a resonant condition that $\delta = 0$. The Lindblad terms $\mathcal{L}_{\mathrm{L(R)}}[\rho]$ describes the dissipations to the left or right of the waveguide, respectively, which is given by
\bea
	\mathcal{L}_{\mathrm{L(R)}}[\rho] =&& \gamma_{\mathrm{L(R)}}
	\sum_{\mu,\nu=1}^{N} e^{\mp i k (x_{\mu} - x_{\nu})}
	\bigg[ \sigma_{\nu} \rho\, \sigma_{\mu}^{\dagger}\nonumber\\
	&&	- \frac{1}{2} \left\{ \sigma_{\mu}^{\dagger} \sigma_{\nu}, \rho \right\} \bigg],
\eea
where $\{ . , .\}$ denotes an anticommutator. 

We define the total decay rate $\gamma = \gamma_L + \gamma_R$ and the directionality factor $D = (\gamma_R - \gamma_L)/\gamma$ \cite{Mitsch2014}. Here, $D = 0$ corresponds to equal leftward and rightward decay rates, characterizing the reciprocal coupling regime, whereas $D = \pm 1$ represents fully unidirectional decay to the right or left, respectively, corresponding to the maximally chiral coupling regime \cite{Gardiner1993, Carmichael1993, Stannigel2012, Kien2017, Buonaiuto2019, Jen2020_disorder, Berman2020, Chen2023, Wang2022, Jen2021_bound, Jen2022}. In the weak-excitation regime, the atomic dynamics can be approximated by restricting the Hilbert space to the ground state and the single-excitation manifold. The system predominantly occupies the collective ground state, with only a small population distributed among singly excited states. Accordingly, the system state can be expanded in terms of the collective ground state $|G\rangle = |g\rangle^{\otimes N}$ and the singly excited states $|j\rangle = |g\rangle^{\otimes(j-1)} \otimes |e\rangle \otimes |g\rangle^{\otimes(N-j)}$, yielding
\begin{equation}
|\Psi(t)\rangle = \frac{1}{\sqrt{1 + \sum^N_{\mu = 1}|p_\mu(t)|^2}}\left(|g\rangle^{\otimes N} + \sum^N_{\mu = 1}p_\mu(t)|\psi_\mu\rangle\right).
\end{equation}
In the above state function, $p_\mu$ denotes the probability amplitude, which satisfies the weak field assumption, $\sum^N_{\mu = 1}|p_\mu(t)|^2 \ll 1$. By solving the system dynamics in the low-energetic sectors in Eq. (\ref{rho}), which reduces to solving it with non-Hermitian Schr\"{o}dinger equations \cite{El-Ganainy2018, Ashida2020, Bergholtz2021}, the evolution of the system becomes
\begin{equation}
	\dot{p}_{\mu} = -i\Omega\, e^{i k x_{\mu} \cos\theta}
	+ \sum_{\nu=1}^{N} M_{\mu\nu}\, p_{\nu},
\end{equation}
with the coupling matrix expressed as 
\begin{equation}
	M_{\mu\nu} =
	\begin{cases}
		-\gamma_{L}\, e^{i k \lvert x_{\mu} - x_{\nu} \rvert}, & \mu < \nu, \\[4pt]
		i\delta_{\mu} - \tfrac{\gamma_{L} + \gamma_{R}}{2}, & \mu = \nu, \\[4pt]
		-\gamma_{R}\, e^{i k \lvert x_{\mu} - x_{\nu} \rvert}, & \mu > \nu.
	\end{cases}
\end{equation}
For steady-state solutions, when $d\rho/ dt = 0$, we obtain 
\begin{equation}
	 \tilde{p}_{\mu} \equiv p_{\mu}(t \to \infty)
	= i\Omega \sum_{\nu=1}^{N} \left( M^{-1} \right)_{\mu\nu}
	e^{i k x_{\nu} \cos\theta}.
\end{equation}
For convenience, we focus on a uniform array with interparticle spacing $d = x_\mu - x_{\mu-1}$, and adopt a dimensionless spacing $\xi=k d$ in our analysis throughout the work. 

\subsection{Trace distance and eigenmodes}

To quantify the distinguishability between quantum states, we employ the trace distance as a time-dependent measure of the system's relaxation toward its steady state \cite{Ares2025, Teza2026}. The trace distance is defined as
\begin{equation}
	T(t) = \frac{1}{2}{\rm Tr}\sqrt{A^\dagger A},
\end{equation}
where $A = \rho(t)-\rho_{ss}$ represents the difference between the time-evolved density matrix $\rho(t)$ and the steady-state density matrix $\rho_{ss}$. In the weak-excitation regime, the system dynamics can be approximated within the Hilbert space spanned by the collective ground state and the singly excited states. Accordingly, the density matrix $\rho(t)$ is represented by an $(N+1)\times(N+1)$ matrix in this truncated Hilbert space.

Starting from the steady state, a local quench is applied at $t=0$ by resetting atom $j$ to its ground state, such that $p_j(0)=0$, while the other amplitudes remain unchanged and the external drive remains continuously applied. To identify the QME, we compare the relaxation dynamics of the entire system following a local quench of atom $A$ with those following a quench of atom $B$. The system exhibits a QME when the initial trace distance satisfies $T_A(0)>T_B(0)$, but the two relaxation curves cross at a later time, such that $T_A(t)<T_B(t)$. Here, $T_A(t)$ and $T_B(t)$ quantify the distances of the respective system states from the common steady state. The QME is classified as persistent if the reversed ordering is maintained at all subsequent times. In contrast, a transient QME occurs when the relaxation curves cross again, restoring their original ordering. Assuming a finite number of genuine crossings, the total number of crossings is therefore odd for a persistent QME and even for a transient QME. The emergence of the QME requires not only a larger initial deviation from the steady state but also sufficiently rapid relaxation over some interval to reverse the ordering of the trace distances.

To investigate the emergence of the QME and elucidate its underlying mechanisms, we perform an eigendecomposition of the interaction matrix $M$. The $N\times N$ matrix $iM$ has $N$ complex eigenvalues $E_n=\omega_n-i\gamma_n$ and their corresponding eigenvectors \cite{Yu2026}. Later in the subsequent sections, we sort the eigenvectors according to the magnitude of the real parts of their eigenvalues, $|\omega_n|$, where $\omega_n$ represents the frequency shift and $\gamma_n$ characterizes the decay rate. An eigenmode is classified as subradiant when $\gamma_n < \gamma$ and superradiant when $\gamma_n > \gamma$. Owing to their distinct decay rates, superradiant modes predominantly contribute to the early-time relaxation dynamics, whereas long-lived subradiant modes become increasingly important at later times as the system approaches its steady state. Moreover, the spatial distributions of these collective eigenmodes can differ significantly, leading to site-dependent relaxation dynamics under different local quenches. By analyzing their decay rates and spatial profiles, we identify the mechanisms governing the emergence of the QME in the following sections.

\begin{figure*}[t]
	\centering
	\includegraphics[width=\linewidth]{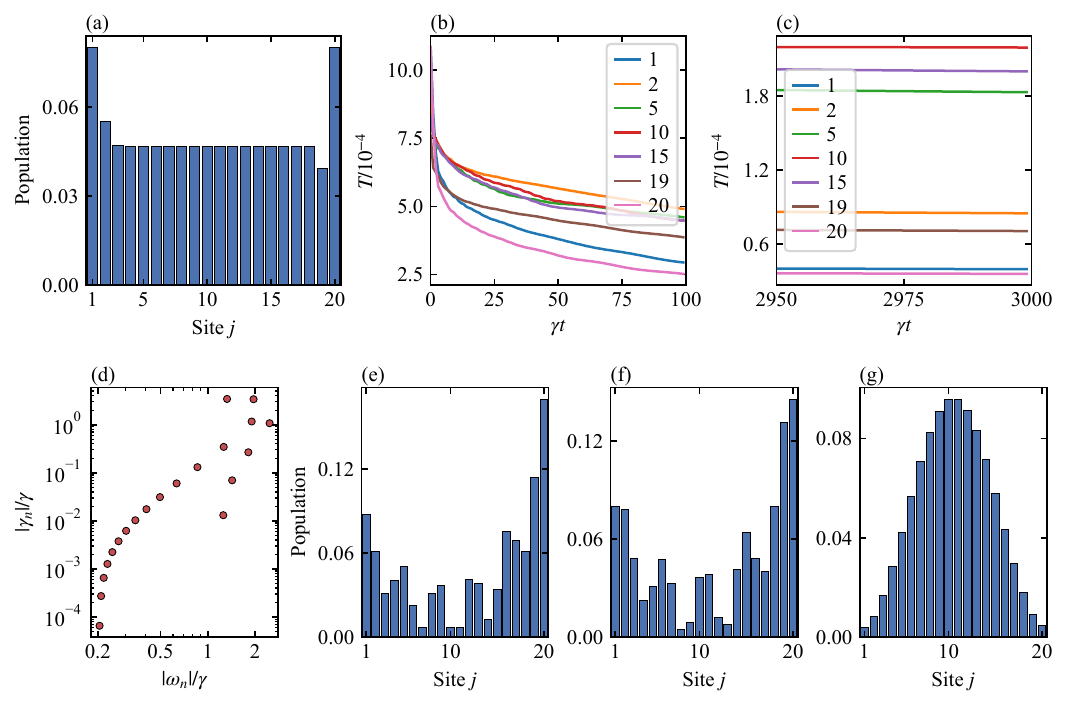}
	\caption{QME in state relaxations toward the bi-edge steady states under local quenches. (a) Steady-state distributions of the bi-edge phase at $N = 20$, $D = 0.1$, and $\xi/\pi = 0.25$. The distribution is concentrated at the edges with the bulk remains constant. (b) Time dynamics of $T$ after single local quenches. Quenches at either atoms at the edges lead to rapid relaxations toward the common steady states. (c) Late time evolution of $T$ in (b), along with multiple crosses from quenched atoms near the edges and the bulk. (d) Distribution of the eigenvalues of matrix $M$. (e-f) The spatial profile for the (e) $15$th and (f) $19$th eigenmode in (d). (g) Normalized spatial distributions for the most subradiant eigenmode in (d).}\label{BE}
\end{figure*}

\section{Edge-enhanced quantum Mpemba effect}
\begin{figure*}[t]
	\centering
	\includegraphics[width=\linewidth]{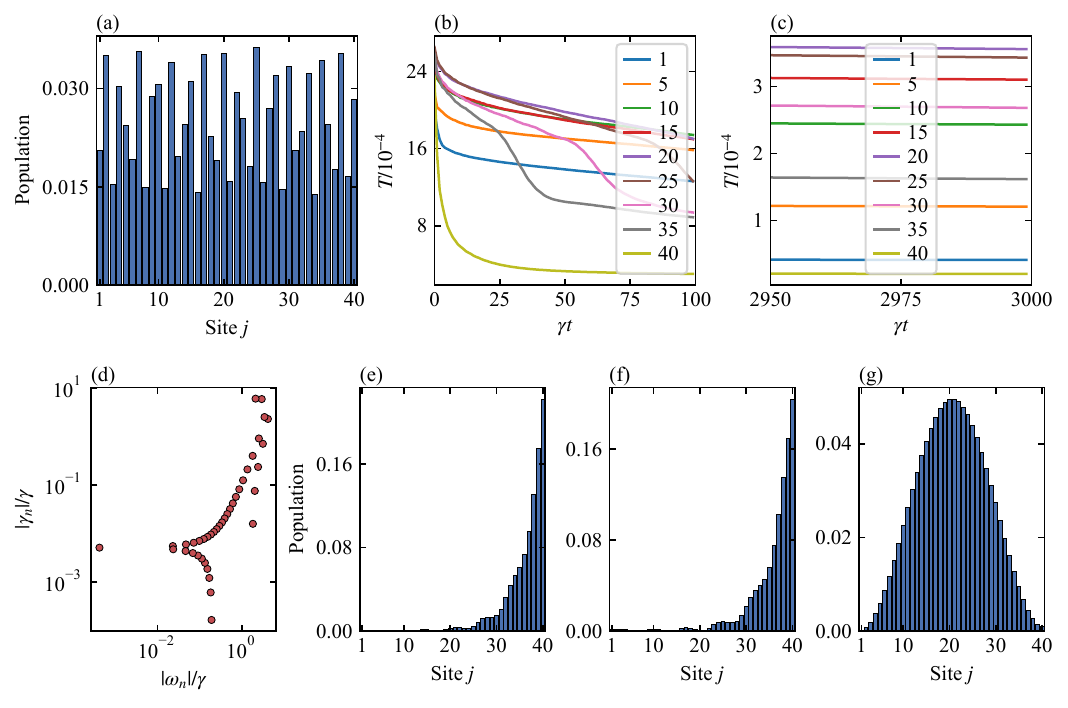}		
	\caption{QME in state relaxations toward the steady states with crystalline orders under local quenches. (a) Steady-state distributions of crystalline phase at $N = 40$, $D = 0.7$, and  $\xi/\pi = 0.15$. The distributions manifest finite periodic structure factors. (b) Time dynamics of $T$ after quenching certain atoms. Quenches at the atoms near both boundaries lead to rapid relaxations toward the common steady states. (c) Late time evolution of $T$ after quenching certain atoms in (b). (d) Distribution of the eigenvalues of matrix $M$. (e-f) The spatial profile for the (e) $34$th and (f) $37$th eigenmode in (d). (g) Normalized spatial distributions for the most subradiant eigenmode in (d).}\label{CO}
\end{figure*}

To illustrate the QME in an atom-waveguide coupled platform, we first consider the bi-edge (BE) steady-state phase in Fig. \ref{BE}(a) with parameters $D = 0.1$ and $\xi/\pi = 0.25$ \cite{Jen2020_steady}. This special steady-state profile near the reciprocal coupling regime $D=0$ showcases pronounced peaks at both edges instead of uniform distributions, which emerges due to the interplay among photon-mediated dipole-dipole interactions determined by $\xi$ and directionality factors $D$.

Since the two edge atoms exhibit higher excitation probabilities than the bulk atoms in the steady state, quenching an edge atom to its ground state induces a larger initial trace distance than quenching an atom in the bulk. The time evolution of the trace distance $T$ following local quenches at different atomic sites is shown in Figs. \ref{BE}(b) and \ref{BE}(c). Quenching either edge atom initially drives the system farther from its steady state, yet the resulting trace distances rapidly decrease below those associated with quenches in the bulk. As shown in Fig. \ref{BE}(b), the systems with either edge atom quenched maintain the smallest trace distances after $\gamma t = 20$, demonstrating a persistent QME driven by rapid early-time relaxation. Among the remaining atomic sites, quenching the atom at site $j=2$ from the edge produces the largest initial trace distance. Interestingly, its relaxation trajectory exhibits no crossings with the other trace-distance curves before $\gamma t = 100$, but undergoes multiple crossings at later times as the system approaches its steady state, as shown in Fig. \ref{BE}(c). These observations reveal a persistent QME, accompanied by multiple crossings of relaxation trajectories, highlighting the rich spatial dependence of relaxation dynamics on the location of the quenched atom. 

To understand the origin of this edge-enhanced relaxation in Fig. \ref{BE}(b), we examine the spatial profiles of the dominant dynamical modes. As shown in Fig. \ref{BE}(d), we sort the eigenvectors based on the magnitude of the real parts of the eigenvalues $|\omega_n|$ with associated eigen-decay rates $\gamma_n$. The spatial distributions of the two most superradiant eigenmodes are shown in Figs. \ref{BE}(e) and \ref{BE}(f). Both show localized distributions near the boundaries, with a slightly stronger localization on the right edge due to the directionality $D>0$. These spatial profiles suggest that quenches at the edges have a stronger overlap with the rapidly decaying superradiant modes, thereby accelerating the initial relaxation compared with quenches in the bulk. However, for positive directionality ($D>0$ or $\gamma_R>\gamma_L$), the initial relaxation dynamics exhibit a pronounced spatial asymmetry that cannot be fully explained by the mode overlap alone. In particular, although the 19th mode has a larger amplitude at site $j=2$ than at most bulk sites, the initial relaxation speed following a quench at site $j=2$ remains comparable to that of bulk atoms and is substantially slower than that following a quench at site $j=1$. This observation highlights the important role of boundary geometry in determining the relaxation dynamics. An edge atom has neighboring atoms on only one side, allowing photons emitted toward the exterior of the array to escape directly into the waveguide without further interacting with other atoms. By contrast, the atom at site $j=2$ has neighboring atoms on both sides, where waveguide-mediated interactions and photon re-scattering can modify the decay dynamics and suppress the initial relaxation speed. Therefore, the rapid relaxation following an edge-atom quench arises from the combined effects of strong coupling to superradiant modes and direct photon emission into the waveguide. Furthermore, the preferential rightward emission under $D>0$ enhances the spatial asymmetry, leading to faster relaxation following a quench at the right edge than at the left edge.

Notably, the relaxation dynamics change substantially at late times. As shown in Fig. \ref{BE}(c), the trace distances become increasingly ordered according to the atoms' distances from the physical boundaries. Approaching $\gamma t=3000$, the trace distances are the smallest for the cases with quenched edge atoms and the largest for quenched atoms located around the center of the array, indicating that resetting operations around the center of the array makes the system decay more slowly than those near the edges. This late-time behavior can be understood from the underlying eigenmode structure. As the rapidly decaying modes become progressively suppressed, the dynamics are increasingly governed by the most subradiant eigenmode. In Fig. \ref{BE}(g), we plot this slowest-decaying mode which has a bell-shaped spatial distribution concentrated around the center of the array. Consequently, a quench near the center has a stronger overlap with this long-lived mode, causing the system to sustain longer due to central quenches. 

Taken together, these results show that the early- and late-time relaxation regimes are governed by different physical mechanisms: superradiant modes and boundary-assisted photon emission drive the rapid initial relaxation at the edges, whereas the spatial profile of the most subradiant mode determines the late-time relaxation near the center.

\section{Skin-enhanced quantum Mpemba effect}

We next investigate the QME in a steady-state phase exhibiting crystalline order (CO) for an atomic array with $D=0.7$ and $\xi/\pi=0.15$ \cite{Jen2020_steady}. In this parameter regime, the steady state is characterized by finite structure factors, indicating the emergence of spatially ordered excitation patterns. As shown in Fig. \ref{CO}(a), this CO phase emerges under relatively strong directional decay, where nonreciprocal waveguide-mediated dipole-dipole interactions support periodic spatial modulations. In contrast to the BE phase discussed in Sec.~III, the CO phase exhibits alternating regions of high and low excitation probabilities throughout the bulk of the array.

Despite these distinct steady-state distributions, quenching the rightmost edge atom induces substantially faster initial relaxation of the system than quenching atoms at other sites, as shown in Fig. \ref{CO}(b). Similar to the BE phase, the corresponding trace distance $T$ decreases rapidly at early times and crosses below those associated with quenches at the leftmost edge and at site $j=5$, for example. Consequently, quenching the rightmost edge atom makes the entire system approach its steady state more rapidly, maintaining the smallest trace distance even at later times, as shown in Fig. \ref{CO}(c). This behavior demonstrates a persistent QME, analogous to that observed for the rightmost edge quench in Fig. \ref{BE}(c).

Similar to the discussion in Sec.~III, the rapid initial relaxation can be understood from the contributions of superradiant eigenmodes shown in Fig. \ref{CO}(d) and direct photon emission at the array boundary. As shown in Figs. \ref{CO}(e) and \ref{CO}(f), both superradiant modes are predominantly concentrated in the right half of the array, with particularly large amplitudes near the right boundary. Consequently, quenching the rightmost edge atom strongly excites these rapidly decaying modes, leading to substantially faster initial relaxation than quenches at other sites. The strong directionality ($D=0.7$), corresponding to $\gamma_R\gg\gamma_L$, further facilitates rapid relaxation through preferential rightward photon emission directly into the outgoing waveguide channel. Notably, the pronounced accumulation of superradiant eigenmodes near the right boundary resembles the boundary localization associated with the non-Hermitian skin effect, which becomes particularly prominent as the system approaches the fully chiral coupling regime ($D\rightarrow 1$). Such non-Hermitian localization can lead to anomalous boundary phenomena and modifications of the conventional bulk-boundary correspondence in Hermitian lattices \cite{Lee2016, Yao2018}. We therefore refer to this enhanced boundary-induced relaxation as the skin-enhanced QME, highlighting the interplay between directional dissipation, spatially localized superradiant modes, and accelerated relaxation toward the steady state.

In addition to this skin-enhanced persistent QME, the relaxation dynamics exhibit delayed crossings of trace-distance trajectories, revealing transient QME following quenches away from the right boundary. In particular, pronounced decreases in $T$ occur at approximately $\gamma t=25$, $50$, and $75$ for quenches at sites $j=35$, $30$, and $25$, respectively. This sequential acceleration of relaxation at progressively more interior sites leads to multiple crossings among the trace-distance trajectories. For example, although quenching the atom at $j=35$ initially makes the system relax faster than quenching the atom at $j=1$, its trace distance eventually becomes larger at later times, as shown in Fig. \ref{CO}(c). Such a reversal in the relative ordering of the trace distances illustrates the transient nature of the QME.

At late times, the relaxation dynamics recover the spatial dependence observed in Sec.~III, with smaller trace distances associated with quenches near the physical boundaries and larger ones near the center of the array. Furthermore, the trace distance following a rightmost edge quench remains smaller than that following a leftmost edge quench, consistent with the preferential rightward emission for $D>0$. This late-time behavior can be understood from the most subradiant eigenmode identified in Fig. \ref{CO}(d), whose spatial profile is shown in Fig. \ref{CO}(g). The mode exhibits a bell-shaped distribution concentrated near the center of the array. Consequently, local quenches in the central region couple more strongly to this long-lived mode, resulting in slower relaxation than quenches near either boundary. The ordering of the late-time trace distances in Fig. \ref{CO}(c) can therefore be qualitatively understood from the spatial amplitudes of the most subradiant mode in Fig. \ref{CO}(g), with stronger mode participation generally corresponding to slower relaxation toward the steady state.
 
 \begin{figure*}[t]
 	\centering
 	\includegraphics[width=\linewidth]{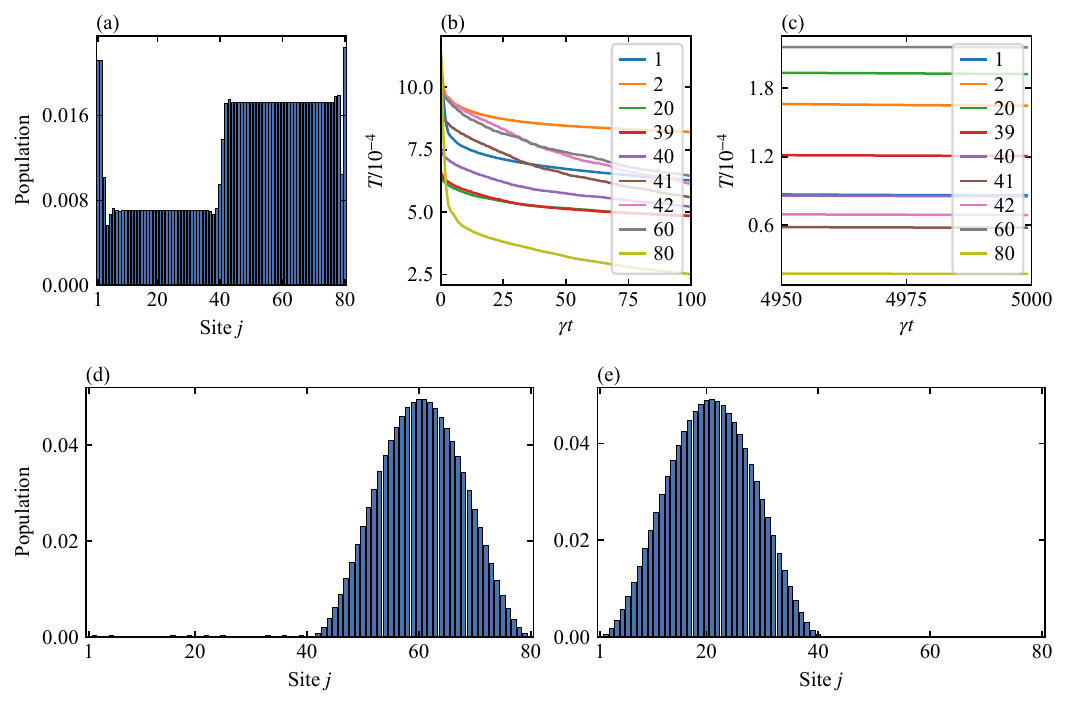}
 	\caption{QME in a dissimilar atomic array with physical boundaries and an internal interface. (a) Steady-state distributions of two arrays with BE phase are coupled together at $D=0.4$, with $\xi_1/\pi=0.2$ and $\xi_2/\pi=0.3$, respectively. Each subarray consists of $40$ atoms with a total $N=80$. The inter-array distance between two subarrays is chosen as the average of $\xi_1$ and $\xi_2$. (b) Time dynamics of $T$ after quenching certain atoms at specific sites. (c) Late time evolution of $T$ after quenching atoms in (b). Normalized spatial distributions of (d) the most subradiant and (e) the second most subradiant eigenmodes.}
 	\label{BE2}
 \end{figure*}

\section{Interface-induced quantum Mpemba effect}

Finally, we investigate the influence of physical boundaries and an internal interface on the QME by considering a composite atomic array formed by coupling two subarrays with different interparticle spacings $\xi_1$ and $\xi_2$ \cite{Chung2024, Handayana2024, Yu2026}. As illustrated in Fig. \ref{BE2}, the system consists of $N=80$ atoms, with the interface located between sites $j=40$ and $j=41$, in addition to the two physical boundaries at the ends of the entire array. The steady-state excitation profile in Fig. \ref{BE2}(a) exhibits two coupled BE subarrays. The relaxation dynamics following local quenches at selected atomic sites are presented in Figs. \ref{BE2}(b) and \ref{BE2}(c). Among these sites, quenching the rightmost atom induces the fastest initial relaxation of the system. Since the two coupled subarrays effectively constitute a single atomic array, this rapid relaxation can be attributed to the localization of superradiant modes and preferential directional photon emission at the physical boundary, as discussed in Secs.~III and IV.

A distinct relaxation behavior emerges when comparing local quenches near the interface with those in the bulk of the subarrays. As shown in Fig. \ref{BE2}(b), the trace-distance trajectory following a quench at site $j=39$, near the interface, nearly coincides with that following a quench at site $j=20$ in the left bulk at early times. At later times, however, the system quenched at $j=20$ remains relatively far from the steady state, whereas the system quenched at $j=39$ approaches it more rapidly, as shown in Fig. \ref{BE2}(c). Since the quench at $j=39$ produces a slightly larger initial trace distance, the subsequent crossing of these relaxation trajectories demonstrates an interface-induced QME.

The underlying mechanism can be understood from the spatial distributions of the long-lived subradiant eigenmodes. As shown in Figs. \ref{BE2}(d) and \ref{BE2}(e), the two most subradiant modes exhibit bell-shaped profiles localized near the centers of the left and right subarrays, respectively. Consequently, the interface region has relatively weak participation in both long-lived modes, suppressing their contributions to the late-time relaxation dynamics following an interface quench. This allows the system quenched near the interface to approach its steady state faster than those quenched in the bulk regions. By contrast, quenches at the physical boundaries induce rapid initial relaxation through strong coupling to superradiant modes and direct photon emission into the waveguide, while their weak participation in long-lived subradiant modes also favors faster relaxation at later times. As a result, quenches at the physical boundaries generally produce the fastest overall relaxation, whereas quenches near the interface lead to faster late-time relaxation than those in the bulk, despite their comparatively slower initial dynamics.

These results demonstrate that open-system relaxation is governed by the spatial interplay between superradiant and subradiant eigenmodes. Superradiant modes facilitate rapid initial relaxation following quenches at the physical boundaries, whereas subradiant modes prolong relaxation following quenches in the bulk regions of the two subarrays. Coupling two distinct arrays introduces an additional spatial domain—the internal interface—which is neither strongly influenced by boundary-localized superradiant modes at early times nor significantly coupled to long-lived subradiant modes at late times. Consequently, quenching an atom near the interface induces relatively slow initial relaxation of the system but allows it to approach the steady state more rapidly at later times. This distinct relaxation behavior highlights the interface-induced QME as an alternative mechanism for accelerated relaxation beyond the boundary-enhanced effects discussed in the preceding sections.

\section{Conclusion and Discussion}
\begin{figure}[t]
	\centering
	\includegraphics[width=\linewidth]{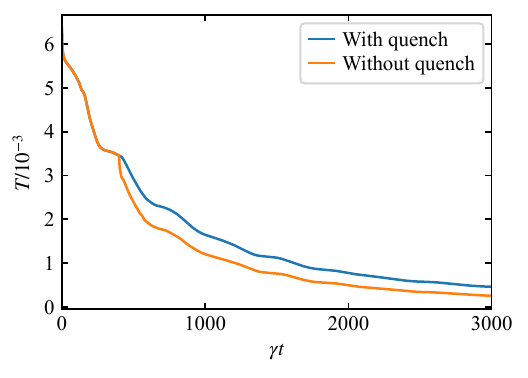}
	\caption{Time evolution with and without a single quench of $j=17$ atom at $\gamma t = 400$ for the array of $N=20$, $D=0.5$, and $\xi/\pi=0.15$.}
	\label{protocol}
\end{figure}
Our results suggest potential applications in accelerating steady-state preparation and controlling state recovery in noisy quantum environments \cite{Yu2026_2}. For example, when an atom is unintentionally reset to its ground state by environmental noise, the subsequent recovery dynamics of the entire system depend strongly on the location of the affected atom. Our simulations reveal that the system generally recovers fastest following a quench at the array boundaries and slowest following a quench near the center. Thus, the spatial location of a noise-induced excitation loss can significantly influence the nonequilibrium relaxation dynamics and the timescale required for the system to recover its steady state.

Another potential application is the acceleration of steady-state preparation through a strategically timed local quench or transient resetting \cite{Solanki2026}. As demonstrated in Fig. \ref{protocol}, we compare the relaxation dynamics with and without a local quench applied at site $j=17$ at $\gamma t=400$. Immediately after the quench, the trace distance decreases and remains below that of the unquenched evolution at subsequent times. This result demonstrates that a suitably timed local quench can accelerate the convergence of the entire system toward its steady state. Although further optimization requires careful selection of both the quenched site and the quench time, this approach suggests a potential route toward controlled two-step Pontus–Mpemba protocols in driven-dissipative quantum systems \cite{Nava2025}. The QME has already been experimentally demonstrated in various platforms, including trapped ions \cite{Joshi2024, Zhang2025}, photonic simulators \cite{Zhan2026}, and superconducting quantum processors \cite{Xu2026}. These advances motivate further exploration of the directional and mode-selective QME predicted here in waveguide QED platforms, including cold-atom arrays, superconducting qubits, and quantum dots.

In conclusion, we have investigated the QME in a weakly driven-dissipative atomic array coupled to a one-dimensional chiral waveguide. We find that early-time relaxation is strongly influenced by boundary-localized superradiant modes and direct photon emission into the waveguide, giving rise to an edge-enhanced QME. Strong decay directionality further enhances this effect by promoting the spatial accumulation of superradiant modes near the preferred boundary. The interplay between rapidly decaying superradiant modes and slower relaxation channels leads to transient QME characterized by multiple trace-distance crossings. At later times, the relaxation dynamics become increasingly dominated by long-lived subradiant modes localized in the bulk of the array. By coupling two distinct subarrays, we further demonstrate an interface-induced QME, where local quenches near the internal interface produce relatively slow initial relaxation but faster late-time relaxation than quenches in the bulk, owing to weak participation in both superradiant and long-lived subradiant modes. Our findings reveal how the spatial structure of collective decay modes and directional dissipation govern anomalous quantum relaxation, offering potential strategies for accelerating steady-state preparation and controlling nonequilibrium dynamics in open quantum systems.
 
\section*{Acknowledgments}
We acknowledge support from the National Science and Technology Council (NSTC), Taiwan, under the Grants No. NSTC-115-2112-M-001035-MY3 and No. NSTC-115-2119-M-001-009, and from Academia Sinica under Grant AS-CDA-113-M04. We are also grateful for support from TG 1.2 of NCTS at NTU.

\section*{Data Availability}
The data that support the findings of this article are not publicly available. The data are available from the authors upon reasonable request.
%%%%%%%%%%%%%%%%%%%%%%%%%%%%%%%%%%%

\end{document}